\documentclass{article} 
\usepackage{epsfig} 
\usepackage{cite}  
\def\NF{N_F}
\def\Poles{{\cal P}oles}
\def\Finite{{\cal F}inite}
\def\Re{\mbox{Re}}
\def\bom#1{{\mbox{\boldmath $#1$}}}

\def\e{\epsilon}
\def\d{\hbox{d}}

\begin{document} 
\unitlength1cm 
\begin{titlepage} 
\vspace*{-1cm} 
\begin{flushright} 
ZU--TH 02/05\\
IPPP/05/03\\
DCPT/05/06\\
hep-ph/0502110\\
February 2005
\end{flushright} 

\vskip 2.5cm

\begin{center} 
{\Large\bf Gluon-Gluon Antenna Functions from Higgs Boson Decay}
\vskip 1.cm 
{\large  A.~Gehrmann--De Ridder}$^{a}$, {\large  T.~Gehrmann}$^{b}$ 
and {\large E.W.N.~Glover}$^{c}$ 
\vskip .7cm 
{\it $^a$ Institute for Theoretical Physics, ETH, CH-8093 Z\"urich,
Switzerland} 
\vskip .4cm 
{\it $^b$ Institut f\"ur Theoretische Physik, Universit\"at Z\"urich,
Winterthurerstrasse 190,\\ CH-8057 Z\"urich, Switzerland} 
\vskip .4cm 
{\it $^c$ Institute for Particle Physics Phenomenology, University of Durham,
South Road,\\ Durham DH1 3LE, England} 
\end{center} 
\vskip 2.6cm 
\begin{abstract} 
Antenna functions describe the infrared singular behaviour of 
colour-ordered
QCD matrix elements due to the emission of unresolved partons inside 
an antenna formed by two hard partons. 
In this paper, we show that antenna functions for hard 
gluon-gluon  pairs can be systematically 
derived from the effective Lagrangian describing Higgs boson decay into 
gluons, and compute the 
the infrared structure of the colour-ordered 
Higgs boson decay matrix 
elements at NLO and NNLO. 
\end{abstract} 
\vfill 

\end{titlepage} 

\newpage 

\renewcommand{\theequation}{\mbox{\arabic{section}.\arabic{equation}}}

\section{Introduction}
\setcounter{equation}{0}
Perturbative QCD corrections to exclusive jet observables are at present 
restricted to the next-to-leading order in perturbation theory, which 
is often insufficient to match the experimental precision on jet production 
reactions~\cite{dissertori}. In the recent past, much progress was made to 
extend these calculations to the next-to-next-to-leading 
order (NNLO) in perturbation 
theory~\cite{mvv,twol,twolmeth,nlomult,cullen,onelstr,onel,campbell,
campbellandother,nnlosub,secdec,ggh,our2j}
and first results for exclusive NNLO cross sections became available 
recently~\cite{babis2j,babishiggs,our3j}. 

The calculation of NNLO corrections to jet observables requires a 
method for the extraction of real radiation singularities arising 
from the emission of up to two unresolved (soft or collinear) partons in 
the final state.  Several methods have been 
proposed recently~\cite{nnlosub} to accomplish the task of 
constructing so-called NNLO subtraction terms. 

In~\cite{our2j}, we described the derivation of NNLO subtraction terms 
for $e^+e^- \to 2j$ based on full four-parton tree-level 
and three-parton one-loop matrix elements, which can be integrated
analytically over the appropriate phase spaces~\cite{ggh}. 
These NNLO subtraction terms were used subsequently~\cite{our3j} 
in the calculation of the 
NNLO corrections to one of the colour factors contributing to 
$e^+e^- \to 3j$.

Subtraction terms derived from full matrix elements can be viewed as 
antenna functions, encapsulating all singular limits due to
unresolved  
partonic emission between two colour-connected hard
partons~\cite{cullen,ant}. 
In particular, process-independent 
antenna functions describing 
arbitrary QCD multiparticle processes can be directly related to three-parton 
matrix elements at NLO (one unresolved parton radiating between two 
colour-connected hard partons)
and four-parton matrix elements at NNLO (two unresolved 
partons radiating between  two 
colour-connected hard partons).

QCD calculations of jet observables require three different types of 
antenna functions, corresponding to the different pairs of hard partons 
forming the antenna: quark-antiquark, quark-gluon and gluon-gluon antenna 
functions. The quark-antiquark antenna functions can be obtained from 
the $e^+e^- \to 2j$ real radiation corrections at NLO and NNLO~\cite{our2j}. 
In~\cite{chi}, we described how the quark-gluon antenna functions 
could be derived from the purely QCD 
(i.e.\ non-supersymmetric) NLO and NNLO corrections to the decay of 
a heavy neutralino into a massless gluino plus partons. It is the purpose 
of this letter to complete the derivation of NLO and NNLO antenna functions
by considering the corrections to the 
decay of a Higgs boson into gluons as template for the gluon-gluon antenna 
functions. 

The Higgs boson
coupling to gluons is mediated through massive quark loops, which 
decouple for large quark masses, thus yielding an effective theory 
containing the interaction of the Higgs field with the gluonic field strength 
tensor~\cite{hgg}. In this effective theory, the Higgs 
boson decay rate~\cite{kniehl}
and inclusive 
Higgs boson production cross sections~\cite{higgsprod1,higgsprod2,higgsprod3}
were computed to NNLO. Most recently, NNLO results for the exclusive Higgs 
boson production cross section~\cite{babishiggs} were obtained as well. 

In the following, we will show which individual real radiation processes 
contribute to the Higgs boson
decay in the effective theory at NLO and NNLO, and 
that the real radiation singularities arising at these orders precisely 
match the infrared singularity structure obtained from an infrared 
factorization formula~\cite{catani}, such that these Higgs boson decay matrix 
elements can be used to derive the gluon-gluon antenna functions at NNLO.

\section{Effective Lagrangian and Feynman rules}
\setcounter{equation}{0}

At tree level, the Higgs boson does not couple either to the gluon or to massless 
quarks. In higher orders in perturbation theory, heavy quark loops introduce
a coupling between the Higgs boson and gluons. In the limit of infinitely 
massive quarks, these loops give rise to an effective Lagrangian~\cite{hgg} 
mediating the 
coupling between the scalar Higgs field and the gluon field strength tensor:
\begin{equation}
{\cal L}_{{\rm int}} = -\frac{\lambda}{4} H F_a^{\mu\nu} F_{a,\mu\nu}\ .
\label{eq:lagr}
\end{equation}
The coupling $\lambda$ has inverse mass dimension. It can be computed 
by matching~\cite{kniehl,kniehl2} 
the effective theory to the full standard 
model cross sections~\cite{higgsfull}.

The Feynman rules following from this Lagrangian are:
\begin{eqnarray}
\parbox{4cm}{\epsfig{file=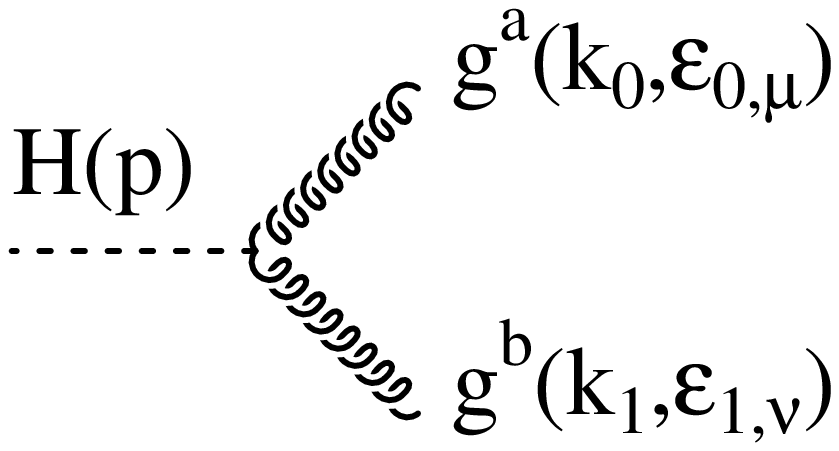,width=4.8cm}} &=&i  \lambda
\delta^{ab} \left( g^{\mu\nu} k_0\cdot k_1 - k_0^{\nu} k_1^{\mu}  \right) \;,
\\[6mm]
\parbox{4cm}{\epsfig{file=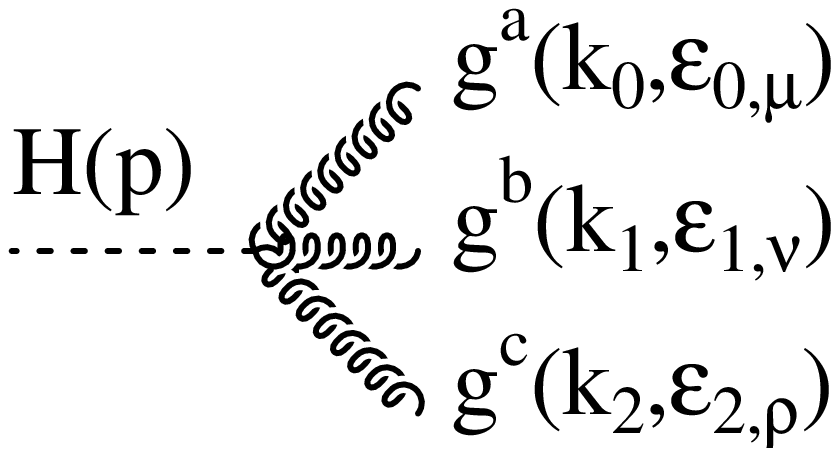,width=4.8cm}} &=& - g_s \lambda
f^{abc} \left( g^{\mu\nu} \left(k_0^{\rho} - k_1^{\rho} \right)
              +g^{\nu\rho} \left(k_1^{\mu} - k_2^{\mu} \right)
              +g^{\rho\mu} \left(k_2^{\nu} - k_0^{\nu} \right)
 \right)\;, \\[6mm]
\parbox{4cm}{\epsfig{file=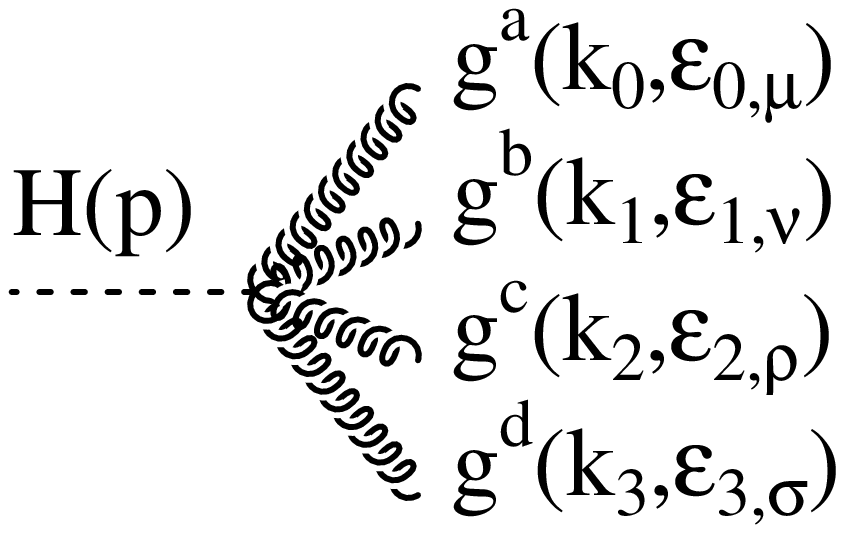,width=4.8cm}} &=& -i g_s^2 \lambda
\bigg[
f^{abe}f^{cde}\left(g^{\mu\rho}g^{\nu\sigma}-g^{\mu\sigma}g^{\nu\rho} \right)
\nonumber \\[-1cm] &&
+f^{ade}f^{bce}\left(g^{\mu\nu}g^{\rho\sigma}-g^{\mu\rho}g^{\nu\sigma} \right)
+f^{ace}f^{dbe}\left(g^{\mu\sigma}g^{\nu\rho}-g^{\mu\nu}g^{\rho\sigma}\right)
\bigg].
\end{eqnarray}
The
momenta are always incoming.

In the present context, the value of $\lambda$ is irrelevant, but we 
do have to take into account that $\lambda$ is renormalized in the 
effective theory. 
The renormalization constant of $\lambda$ 
was computed to all orders in~\cite{condens}; it reads
\begin{equation}
Z_\lambda = \frac{1}{1-\beta(\alpha_s)/\e} = 
1 - \frac{\alpha_s}{2\pi} \, \frac{\beta_0}{\e}
 + \left(\frac{\alpha_s}{2\pi}\right)^2 \left[
\frac{\beta_0^2}{\e^2} - \frac{\beta_1}{\e} \right] 
+ {\cal O} (\alpha_s^3)\; ,
\label{eq:zlambda}
\end{equation}
with
\begin{equation}
\beta_0 = \frac{11 N - 2 N_F}{6}\;, \qquad \beta_1 =
 \frac{34 N^3 - 13 N^2  N_F+   3 N_F}{12N}\;.
\label{eq:qcdbeta}
\end{equation}

\section{Colour-ordered amplitudes in Higgs boson decay}
\setcounter{equation}{0}

\begin{figure}[t]
\begin{center}
\parbox{6cm}{\begin{picture}(6,3)
\thicklines
\put(0.0,1.5){\line(1,1){1.0}}
\put(0.0,1.5){\line(1,-1){1.0}}
\put(0.0,1.4){\line(1,1){1.0}}
\put(0.0,1.4){\line(1,-1){1.0}}
\put(0.0,1.45){\circle*{0.2}}
\put(1.1,0.45){\makebox(0,0)[l]{$a_2$}}
\put(1.1,2.45){\makebox(0,0)[l]{$a_1$}}
\put(1.7,1.45){\makebox(0,0)[l]{$\cdot \delta^{a_1a_2}\; =$}}
\put(4.0,1.45){\line(1,1){1.0}}
\put(4.0,1.45){\line(1,-1){1.0}}
\put(4.1,1.45){\line(1,1){0.9}}
\put(4.1,1.45){\line(1,-1){0.9}}
\put(5.1,0.45){\makebox(0,0)[l]{$a_1$}}
\put(5.1,2.45){\makebox(0,0)[l]{$a_1$}}
\end{picture}}
\end{center}
\caption{Colour flow contained in tree level 
decay $H\to  g g$. Double (single) lines denote adjoint
(fundamental) colour indices.}
\label{fig:ant0}
\end{figure}
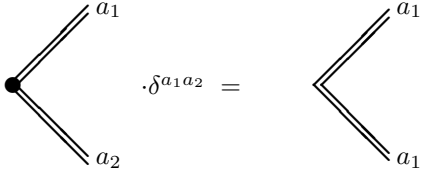

The basic process for the decay of a Higgs boson into 
partons is 
$H(q) \to  g(p_1) g(p_2)$. Its amplitude reads
\begin{equation}
{\cal M}^0_{g_1 g_2} 
= i \lambda \delta^{a_1 a_2} M_{g g}^0 (p_1,p_2) \; .
\label{eq:master0}
\end{equation}
The amplitude contains two 
colour connected (hard) partons which form  two antennae, since
unresolved parton emission can take place on both fundamental colour 
lines connecting the gluons $p_1$ and $p_2$, as illustrated in 
Figure~\ref{fig:ant0}.

The squared matrix element, averaged over identical gluons in the 
final state is
\begin{equation}
{\cal T}^0_{g g}(q^2) \equiv \frac{1}{2}\,|{\cal M}^0_{g_1 g_2}|^2 
= \frac{1}{2}\,\lambda^2 \left(N^2-1\right) 
|M_{g g}^0 (p_1,p_2)|^2  = 
\frac{1}{4}\, \left(N^2-1\right)\, \lambda^2(1-\e)(q^2)^2 \; .
\end{equation}
${\cal T}^0_{g g}(q^2) $  serves as normalization for 
antenna functions obtained from higher order 
corrections to this matrix element.

To demonstrate the cancellation of infrared divergences at NLO, we
compute the renormalized one-loop QCD correction to the 
$H(q) \to g(p_1) g(p_2)$ decay,
\begin{eqnarray}
{\cal T}^1_{g g}(q^2) &\equiv& \frac{1}{2}\,
2\mbox{Re}|{\cal M}^0_{g_1 g_2}{\cal M}^{1,*}_{g_1 g_2}
|\nonumber \\
&=&\left(\frac{\alpha_s}{2\pi}\right)
2 (q^2)^{-\e}\, 
{\cal T}^0_{g g}(q^2) 
\Bigg\{ N \Bigg[ -\frac{1}{\e^2} - \frac{11}{6\e}  + 
\frac{7\pi^2}{12} 
+ \left( -1 + \frac{7}{3}\zeta_3 \right)
\e 
+ \left( - 3
- \frac{73\pi^4}{1440} \right) \e^2  \Bigg] \nonumber \\
&& \hspace{3.5cm}+ 
\frac{N_F}{3\e} + {\cal O}(\e^3)\Bigg\} \;.
\end{eqnarray}
The infrared poles of this one-loop correction can be expressed in terms 
of the infrared singularity operator\cite{catani}
\begin{equation}
\bom{I}_{gg}^{(1)}(\epsilon,q^2)
= - \frac{e^{\epsilon\gamma}}{2\Gamma(1-\epsilon)} \left[
N\,
\left( \frac{1}{\e^2}+\frac{\beta_0}{N\e}\right)
\,\left(-q^2\right)^{-\e}\,\right]\;  
\label{eq:I1}
\end{equation}
as
\begin{equation}
\Poles\left( {\cal T}^1_{g g}(q^2)  \right) = 
\left(\frac{\alpha_s}{2\pi}\right)\;
4 \Re \bom{I}_{gg}^{(1)}(\epsilon,q^2)\;{\cal T}^0_{g g}(q^2)\;.
\label{eq:poles1l}
\end{equation}
This expression has to be compared to the 
$2 \Re \bom{I}_{q\bar q}^{(1)}(\epsilon,q^2)$, which is obtained in the 
decay of a virtual photon into a quark-antiquark pair 
$\gamma^* \to q \bar q$ at one loop~\cite{our2j} 
and the factor
$4 \Re \bom{I}_{q g}^{(1)}(\epsilon,q^2)$, which is obtained in the 
decay of a neutralino into a gluino-gluon pair 
$\tilde{\chi} \to \tilde{g} g$ at one loop~\cite{chi}.
The factor $4$ in 
(\ref{eq:poles1l}) appears since the leading order process 
$H\to g g$ contains two distinct 
gluon-gluon antennae, just as $\tilde{\chi} \to \tilde{g} g$ contains 
two quark-gluon antennae, but in contrast to 
the single quark-antiquark antenna in 
$\gamma^* \to q \bar q$.

\section{NLO antenna functions}
\setcounter{equation}{0}
Two different emissions off a gluon-gluon pair appear at NLO:
either the emission of an additional gluon or the splitting of one
gluon into a quark-antiquark pair. In the context of
Higgs boson decay, these correspond to the tree level 
processes $H \to g gg$ and $H\to g q 
\bar q$. 

The tree level amplitude for $H(q) \to g(p_1) 
g (p_2) g(p_3)$
contains only a single colour structure, $f^{a_1a_2a_3}$: 
\begin{equation}
{\cal M}^0_{g_1 g_2 g_3} 
= i \lambda g \,f^{a_1a_2a_3}\,
 M_{g g g}^0 (p_1,p_2,p_3) \;.
\end{equation}
 Squaring the
matrix element and dividing by a symmetry factor to account for 
identical gluons in the final state yields
\begin{equation}
\frac{1}{3!}\,|{\cal M}^0_{g_1 g_2 g_3}|^2 = \lambda^2 g^2 \,
\left(N^2-1\right) N 
\, \frac{1}{3!} \, |M_{g g g}^0 (p_1,p_2,p_3)|^2 \;,
\end{equation}
with
\begin{eqnarray}
\frac{1}{3!}\,|M_{g g g}^0 (p_1,p_2,p_3)|^2  &=& 
\frac{1}{2}\, (1-\e)\, \frac{1}{3} \Bigg(
\frac{2 s_{123}^2 s_{12}}{ s_{13}  s_{23}}
+\frac{2 s_{123}^2 s_{13}}{ s_{12}  s_{23}}
+\frac{2 s_{123}^2 s_{23}}{ s_{12}  s_{13}}
\nonumber \\ &&\hspace{2cm}
+\frac{2s_{12}s_{13}}{s_{23}}
+\frac{2s_{12}s_{23}}{s_{13}}
+\frac{2s_{13}s_{23}}{s_{12}}
 +12  s_{123}
\Bigg) - \frac{2}{3} s_{123} \; .
\label{eq:gggM}
\end{eqnarray}
The factor $1/3$ in the above equation reflects the fact that the $H\to ggg$
matrix element contains three different antenna configurations
(corresponding to the three different possibilities of identifying 
the two hard gluons and the one unresolved gluon).
The effect of the symmetrization over the 
three gluons is that these three antenna configurations are averaged over. 
To illustrate the antenna factorization, the leading order matrix element 
(without the symmetrization factor for two identical gluons) is factored
out. 

The behaviour of this matrix element in the kinematical limits where one 
parton becomes unresolved is as follows:
\begin{enumerate}
\item Collinear limits:
\begin{eqnarray}
\frac{1}{3!}\,|{\cal M}^0_{g_i g_j g_k}|^2 
&\stackrel{g_i \parallel
g_j}{\longrightarrow}& \left({4\pi\alpha_s}\right)\;
2\,\frac{{\cal T}^0_{g g}(s_{ijk})}{3} \frac{1}{s_{ij}}\,
N \,  P_{g\to gg}(z)\;,
\end{eqnarray}
with $z$ being the momentum fraction of one of the collinear partons and
the splitting function 
\begin{displaymath}
P_{g\to gg}(z) = 2\left[\frac{z}{1-z}+ \frac{1-z}{z}+z(1-z)\right]\;.
\end{displaymath}
\item Soft limits:
\begin{eqnarray}
\frac{1}{3!}\,|{\cal M}^0_{g_i g_j g_k}|^2 &\stackrel{g_j \to
  0}{\longrightarrow}& \left({4\pi\alpha_s}\right)\;
2\,\frac{{\cal T}^0_{g g}(s_{ijk})}{3}\,  N\, 
\frac{2s_{ik}}{s_{ij}s_{jk}}\; . 
\end{eqnarray}
\end{enumerate}
Besides the symmetry factor $1/3$ accounting for the average over the 
three different antenna configurations, we observe an overall factor $2$, 
corresponding to the presence of two distinct antenna functions
in the basic two-parton matrix element, Figure~\ref{fig:ant0}.

To obtain antenna functions describing the emission of an unresolved 
gluon $j$ off an antenna containing two hard gluons $i$, $k$, 
the matrix element 
(\ref{eq:gggM}) has to be split into three individual antenna configurations.
Each individual antenna configuration contains only one soft limit.
Each collinear $g\to gg$ is split between the 
two antenna configurations 
appropriate to the two final state gluons involved in
the splitting, as 
discussed in~\cite{cullen,ant}. 

Integration over the dipole phase space~\cite{ggh} yields 
\begin{eqnarray}
{\cal T}^1_{g gg}(q^2) 
&\equiv&
\int \d \Phi_{D,g gg}\;\frac{1}{3!}|{\cal M}^0_{g_1 g_2 g_3}|^2
\nonumber \\
&=&  \left(\frac{\alpha_s}{2\pi}\right)\; N \;{\cal T}^0_{g g}(q^2)
\left(q^2 \right)^{-\e} 
  \Bigg[
\frac{2}{\e^2} + \frac{11}{3\e} + \frac{73}{6} - 
\frac{7\pi^2}{6} 
+ \left( \frac{451}{12} - \frac{77\pi^2}{36} - \frac{50}{3}\zeta_3 \right)
\e \nonumber \\
&& \hspace{2cm}
+ \left( \frac{2729}{24}
          - \frac{511\pi^2}{72}
          - \frac{275}{9}\zeta_3
          - \frac{71\pi^4}{720} \right) \e^2  + {\cal O}(\e^3) \Bigg] \; .
\end{eqnarray}

The tree level amplitude for $H(q) \to g(p_1)
q(p_3) \bar q(p_4)$
contains only a single colour structure $T^{a_1}_{i_3 i_4}$:
\begin{equation}
{\cal M}^0_{g_1 q_3 \bar q_4} 
= i \lambda g \,T^{a_1}_{i_3 i_4}\, M_{g q \bar q}^0 (p_1,p_3,p_4)
\;,
\end{equation}
yielding 
\begin{equation}
|{\cal M}^0_{g_1 q_3 \bar q_4}|^2 = \lambda^2 
g^2 \frac{N^2-1}{2} 
|M_{g q \bar q}^0 (p_1,p_3,p_4)|^2 \;,
\end{equation}
with 
\begin{equation}
|M_{g q \bar q}^0 (p_1,p_3,p_4)|^2 = 
\frac{1}{2}\,
 \left( 1-\e \right) \left(2 \frac{\left(s_{13}+s_{14}\right)^2}{s_{34}}
 \right) - 2 \frac{s_{13}s_{14}}{s_{34}}\;.
\end{equation}
The only singular configuration contained in this matrix element is 
the collinear quark-antiquark limit, which is as follows:
\begin{equation}
|{\cal M}^0_{g_1 q_3 \bar q_4}|^2 \stackrel{q_3 \parallel
\bar q_4}{\longrightarrow} 2\, \left({4\pi\alpha_s}\right)
\;
{\cal T}^0_{g g}(s_{134}) \frac{1}{s_{34}}\,
  P_{g\to q\bar q}(z) \;,
\end{equation}
with the collinear splitting function
\begin{displaymath}
P_{g\to q\bar q}(z) = 1 - \frac{2z(1-z)}{1-\e}\;.
\end{displaymath}
The factor $2$ arises from the fact that two 
gluon-gluon antennae are contained
in the matrix element (\ref{eq:master0}) as above.

Integration over the dipole phase space~\cite{ggh} and summing 
over final state quark flavours yields 
\begin{eqnarray}
{\cal T}^1_{g q \bar q} (q^2) &\equiv&
\int \d \Phi_{D,g q \bar q}\; \sum_q\, 
|{\cal M}^0_{g_1 q_3 \bar q_4}|^2 
 \nonumber \\
&=&  \left(\frac{\alpha_s}{2\pi}\right)\;
 N_F\, {\cal T}^0_{g g} (q^2) \left(q^2 \right)^{-\e} 
  \Bigg[
- \frac{2}{3\e} -\frac{7}{3}   
+ \left( -\frac{15}{2} + \frac{7\pi^2}{18} \right)
\e \nonumber \\ && \hspace{3cm} 
+ \left( -\frac{93}{4} 
          + \frac{49\pi^2}{36}
          + \frac{50}{9}\zeta_3
\right) \e^2  + {\cal O}(\e^3) \Bigg] \; .
\end{eqnarray}

Summing over both three parton final states, we find
\begin{equation}
\Poles\left( {\cal T}^1_{g g g}(q^2) + 
{\cal T}^1_{g q \bar q}(q^2)  \right) =
-\left(\frac{\alpha_s}{2\pi}\right)\;4 \Re 
\bom{I}_{gg}^{(1)}(\epsilon,q^2)\;{\cal T}^0_{g g}(q^2)\;,
\label{eq:polestree}
\end{equation}
such that the NLO corrected Higgs boson decay rate into  partons 
is finite:
\begin{equation}
\Poles\left( {\cal T}^1_{g g }(q^2) \right) + 
\Poles\left( {\cal T}^1_{g g g}(q^2) + 
{\cal T}^1_{g q \bar q}(q^2)  \right) = 0\;.
\end{equation}
We recover the finite NLO 
contribution to the Higgs boson decay rate into partons in 
the effective theory as
\begin{equation}
\Finite\left( {\cal T}^1_{g g }(q^2) \right) + 
\Finite\left( {\cal T}^1_{g g g}(q^2) + 
{\cal T}^1_{g q \bar q}(q^2)  \right) = \frac{\alpha_s}{2\pi} \left(
\frac{73}{6} N - \frac{7}{3} N_F \right)\, {\cal T}^0_{g g}(q^2)\;,
\end{equation}
which is in agreement with~\cite{kniehl}.

\section{Structure of NNLO antenna functions}
\setcounter{equation}{0}

In the NNLO calculation of jet observables, two different types of antenna 
functions are required: (a) the one-loop correction to the 
three-parton antenna 
functions which appeared at NLO in tree-level form, and (b) 
the tree-level 
four-parton antenna functions. In this section, we present all Higgs
boson 
decay matrix elements needed for the derivation of these antenna functions, 
and demonstrate that these matrix elements contain the same  infrared 
singularities as processes involving final state emission off a gluon-gluon 
antenna. 

The renormalized 
one-loop corrections to the three-parton antenna functions have the 
same colour structure as their tree level counterparts listed above. 
To expose the infrared structure of the resulting one-loop matrix elements, 
they are 
integrated over the corresponding dipole phase space~\cite{ggh}, 
yielding
\begin{eqnarray}
{\cal T}^2_{g gg}(q^2) 
&\equiv&
\int \d \Phi_{D,g gg}\;\frac{1}{6}\,
2 \mbox{Re}\left({\cal M}^0_{g_1 g_2 g_3}
{\cal M}^{1,*}_{g_1 g_2 g_3}\right)
\nonumber \\
&=& \left(\frac{\alpha_s}{2\pi}\right)^2\; {\cal T}^0_{g g}(q^2)
\left(q^2 \right)^{-2\e} 
  \Bigg[ N^2 \Bigg( 
-\frac{9}{2\e^4}  - \frac{121}{6\e^3} + \frac{1}{\e^2} 
\left(-\frac{170}{3} + \frac{71\pi^2}{12} \right) \nonumber \\
&&+ \frac{1}{\e} \left(
 - \frac{23195}{108}
          + \frac{341\pi^2}{18}
          + 72\zeta_3 \right) 
+ \left(  - \frac{173249}{216}
          + \frac{13831\pi^2}{216}
          + \frac{2266}{9}\zeta_3
          - \frac{995\pi^4}{720} 
\right)\Bigg) \nonumber \\
&& + N N_F \Bigg( \frac{2}{\e^3} + \frac{11}{3\e^2} 
+ \frac{1}{\e} \left(\frac{37}{3}
          - \frac{7\pi^2}{6}
\right) 
+ \left(\frac{467}{12}
          - \frac{77\pi^2}{36}
          - \frac{50}{3}\zeta_3
\right)
\Bigg)
+ {\cal O}(\e) \Bigg] \; , \\
{\cal T}^2_{g q\bar q}(q^2) 
&\equiv&
\int \d \Phi_{D,g q\bar q}\;
2 \mbox{Re}\left({\cal M}^0_{g_1 q_3 \bar q_4}
{\cal M}^{1,*}_{g_1 q_3 \bar q_4}\right)
\nonumber \\
&=&  \left(\frac{\alpha_s}{2\pi}\right)^2\;{\cal T}^0_{g g}(q^2)
\left(q^2 \right)^{-2\e} 
  \Bigg[ N N_F \Bigg( 
\frac{4}{3\e^3}  + \frac{25}{3\e^2} + \frac{1}{\e} 
\left(\frac{805}{27} - \frac{16\pi^2}{9} \right)
\nonumber \\
&&+  \left(
  \frac{2926}{27}
          - \frac{947\pi^2}{108}
          - \frac{188}{9}\zeta_3 \right) 
\Bigg) \nonumber \\
&& + \frac{N_F}{N} \Bigg( -\frac{1}{3\e^3} - \frac{41}{18\e^2} 
+ \frac{1}{\e} \left(-\frac{325}{27}
          + \frac{\pi^2}{2}
\right) + \left(-\frac{18457}{324}
          + \frac{41\pi^2}{12}
          + \frac{74}{9}\zeta_3
\right)
\Bigg)\nonumber \\
&& + N_F^2 \Bigg( -\frac{4}{9\e^2} -\frac{7}{9\e} + \left(\frac{85}{162}
          + \frac{\pi^2}{18}
\right)
\Bigg)
+ {\cal O}(\e) \Bigg] \; .
\end{eqnarray}

Three different four-parton final states appear in the gluon-gluon antenna 
functions at NNLO: $gggg$, 
$q\bar q gg$ and $q\bar q q'\bar q'$. In contrast to the 
tree level three-parton Higgs boson decay matrix elements, 
which contained only one non-trivial colour 
ordering each, these four-parton matrix elements all contain several 
colour-orderings.

The amplitude for
$H(q) \to g(p_1) g (p_2) g(p_3) g(p_4)$
 can then be expressed as sum over the permutations 
of the gluon colour indices:
\begin{equation}
{\cal M}^0_{g_1 g_2 g_3 g_4} 
= i \lambda g^4 \sum_{(i,j,k) \in P(2,3,4)} 
\mbox{Tr}(T^{a_1}T^{a_i}T^{a_j}T^{a_k})
M_{g g g g}^0 (p_1,p_i,p_j,p_k) \;,
\end{equation}
where the sum runs over all six permutations of three of the 
gluon colour indices, thus excluding any configurations which can
be related by cyclic permutations of all four colour indices. Its colour 
flow is illustrated in Figure~\ref{fig:f40}.

The resulting squared matrix element, averaged over identical 
final state gluon permutations is 
\begin{eqnarray}
\frac{1}{4!}\,\left|{\cal M}^0_{g_1 g_2 g_3 g_4}\right|^2 
&=& \lambda^2 g^4\, \frac{N^2-1}{16}\, \frac{1}{4!}\,  N^2
\sum_{(i,j,k) \in P(2,3,4)} 
\left|M_{g g g g}^0 (p_1,p_i,p_j,p_k)\right|^2 
\;.
\end{eqnarray}
It should be 
noted that this squared matrix element contains only  the leading colour
term obtained from the squares of the individual colour-ordered amplitudes,
as expected in the colour ordered formulation for a process with four 
gluons~\cite{colord,ddm}. 
\begin{figure}[t]
\begin{center}
\parbox{4cm}{\begin{picture}(4,1.5)
\thicklines
\put(0.25,0){\line(1,0){3.0}}
\put(0.15,-0.3){\line(1,0){3.1}}
\put(0.25,0){\line(0,1){1.0}}
\put(0.15,0){\line(0,1){1.0}}
\put(0.15,0){\line(0,-1){0.3}}
\put(1.15,0){\line(0,1){1.0}}
\put(1.25,0){\line(0,1){1.0}}
\put(2.15,0){\line(0,1){1.0}}
\put(2.25,0){\line(0,1){1.0}}
\put(3.15,0){\line(0,1){1.0}}
\put(3.25,0){\line(0,1){1.0}}
\put(3.25,0){\line(0,-1){0.3}}
\put(0.2,0){\circle*{0.2}}
\put(1.2,0){\circle*{0.2}}
\put(2.2,0){\circle*{0.2}}
\put(3.2,0){\circle*{0.2}}
\put(0.1,1.2){\makebox(0,0)[l]{$a_1$}}
\put(1.1,1.2){\makebox(0,0)[l]{$a_i$}}
\put(2.1,1.2){\makebox(0,0)[l]{$a_j$}}
\put(3.1,1.2){\makebox(0,0)[l]{$a_k$}}
\end{picture}}
\end{center}
\caption{Colour flow contained in the colour ordered amplitude 
$M_{g g g g}^0 (p_1,p_i,p_j,p_k)$ contributing to the 
tree level 
decay $H \to  g g gg$.}
\label{fig:f40}
\begin{center}
\parbox{4cm}{\begin{picture}(4,1.5)
\thicklines
\put(0.15,0){\line(1,0){3.1}}
\put(0.15,0){\line(0,1){1.0}}
\put(1.15,0){\line(0,1){1.0}}
\put(1.25,0){\line(0,1){1.0}}
\put(2.15,0){\line(0,1){1.0}}
\put(2.25,0){\line(0,1){1.0}}
\put(3.25,0){\line(0,1){1.0}}
\put(1.2,0){\circle*{0.2}}
\put(2.2,0){\circle*{0.2}}
\put(0.1,1.2){\makebox(0,0)[l]{$i_3$}}
\put(1.1,1.2){\makebox(0,0)[l]{$a_1$}}
\put(2.1,1.2){\makebox(0,0)[l]{$a_2$}}
\put(3.1,1.2){\makebox(0,0)[l]{$i_4$}}
\end{picture}}
\end{center}
\caption{Colour flow contained in the colour ordered amplitude 
$M_{g g q \bar q }^0 (p_1,p_2,p_3,p_4)$   
contributing to the 
tree level 
decay $H \to  gg q\bar q$.}
\label{fig:g40}
\end{figure}

The tree level amplitude for $H(q) \to g(p_1)
g(p_2) q(p_3) \bar q(p_4)$ contains two colour structures,
\begin{eqnarray}
{\cal M}^0_{g_1 g_2 q_3 \bar q_4} 
&=& i \lambda g^2  \Bigg[ \left(T^{a_1}T^{a_2}\right)_{i_3i_4}
M_{gg q \bar q}^0 (p_1,p_2,p_3,p_4) +
\left(T^{a_2}T^{a_1}\right)_{i_3i_4}
M_{gg q \bar q}^0 (p_2,p_1,p_3,p_4) \Bigg]\;.
\end{eqnarray}

The squared matrix element, averaged over identical gluons in the final state
and summed over quark flavours, 
 reads
\begin{eqnarray}
\frac{1}{2} |{\cal M}^0_{g_1 g_2 q_3 \bar q_4} |^2
 &=& \lambda^2 g^4 \frac{N^2-1}{8} \, N_F
\Bigg\{ N \left[ \left| 
M_{gg q \bar q}^0 (p_1,p_2,p_3,p_4)\right|^2
+  \left| M_{g g q \bar q}^0 
(p_2,p_1,p_3,p_4)\right|^2 \right] 
\nonumber \\ && \hspace{1cm}
- \frac{1}{N} \left|M_{gg q \bar q}^0 (p_1,p_2,p_3,p_4)+
M_{g g q \bar q}^0 
(p_2,p_1,p_3,p_4)
\right|^2 \Bigg\} \;.
\end{eqnarray}

Finally, 
the tree level amplitude for $H (q) \to q(p_1)
q(p_2) q'(p_3) \bar q'(p_4)$ contains only a single colour 
structure, but can contain two flavour structures in the case 
of identical quark flavours $q = q'$:
\begin{eqnarray}
{\cal M}^0_{q_1 q_2 q'_3 \bar q'_4} 
&=& i \lambda g^2  
\Bigg[ T^{a_1}_{i_1i_2} T^{a_1}_{i_3i_4} 
M_{q\bar q q' \bar q'}^0 (p_1,p_2,p_3,p_4) -
\delta_{qq'} T^{a_1}_{i_1i_4} T^{a_1}_{i_3i_2}
M_{q\bar q q' \bar q'}^0 (p_1,p_4,p_3,p_2) \Bigg]\;.
\end{eqnarray}

The squared matrix element, summed over quark flavours and 
averaged over identical configurations becomes
\begin{eqnarray}
\frac{1}{2} |{\cal M}^0_{q_1 q_2 q'_3 \bar q'_4} |^2
 &=& \lambda^2 g^4 \frac{N^2-1}{8} \, \Bigg\{ N_F^2
 \left| 
M_{q \bar q q' \bar q'}^0 (p_1,p_2,p_3,p_4)\right|^2
\nonumber \\
&& \hspace{2cm} - \frac{N_F}{N} \delta_{qq'} \mbox{Re}\left( 
M_{q \bar q q' \bar q'}^0 (p_1,p_2,p_3,p_4)
M_{q \bar q q' \bar q'}^{0,*} (p_1,p_4,p_3,p_2)
\right) \Bigg\} 
 \;.
\end{eqnarray}

The four-parton tree-level Higgs 
boson decay matrix elements can be integrated 
over the tripole phase space~\cite{ggh}, thus making their infrared 
singularity structure explicit,
\begin{eqnarray}
{\cal T}^2_{g ggg}(q^2) 
&\equiv&
\int \d \Phi_{T,g ggg}\;\frac{1}{4!}
\left|{\cal M}^0_{g_1 g_2 g_3 g_4}\right|^2
\nonumber \\
&=& \left(\frac{\alpha_s}{2\pi}\right)^2\; {\cal T}^0_{g g}(q^2)
\left(q^2 \right)^{-2\e} 
  N^2 \Bigg[ 
\frac{5}{2\e^4}  + \frac{121}{12\e^3} + \frac{1}{\e^2} 
\left(\frac{436}{9} - \frac{11\pi^2}{3} \right) \nonumber \\
&&+ \frac{1}{\e} \left(
            \frac{23455}{108}
          - \frac{1067\pi^2}{72}
          - \frac{379}{6}\zeta_3 \right) 
\nonumber \\
&&
+ \left(    \frac{304951}{324}
          - \frac{7781\pi^2}{108}
          - \frac{2288}{9}\zeta_3
          + \frac{479\pi^4}{720} 
\right)
+ {\cal O}(\e) \Bigg] \; , \\
{\cal T}^2_{gg q\bar q }(q^2) 
&\equiv&
\int \d \Phi_{T,gg q\bar q }\;
\frac{1}{2}\,\left|{\cal M}^0_{g_1 g_2 q_3 \bar q_4}\right|^2
\nonumber \\
&=&  \left(\frac{\alpha_s}{2\pi}\right)^2\;{\cal T}^0_{g g}(q^2)
\left(q^2 \right)^{-2\e} 
  \Bigg[ N N_F \Bigg( 
-\frac{3}{2\e^3}  - \frac{155}{18\e^2} + \frac{1}{\e} 
\left(-\frac{523}{12} + \frac{79\pi^2}{36} \right)
\nonumber \\ && \hspace{3.8cm}
+  \left(
  -\frac{16579}{81}
          + \frac{1385\pi^2}{108}
          + 37\zeta_3 \right) 
\Bigg) \nonumber \\
&& + \frac{N_F}{N} \Bigg( \frac{1}{3\e^3} + \frac{41}{18\e^2} 
+ \frac{1}{\e} \left( \frac{1327}{108}
          - \frac{\pi^2}{2}
\right) + \left(\frac{4864}{81}
          - \frac{41\pi^2}{12}
          - \frac{86}{9}\zeta_3
\right)
\Bigg)
+ {\cal O}(\e) \Bigg] \; ,\\
{\cal T}^2_{q\bar q q' \bar q'}(q^2) 
&\equiv&
\int \d \Phi_{T,q\bar qq'\bar q'}\;
\frac{1}{2}\,\left|{\cal M}^0_{q_1 \bar q_2 q'_3 \bar q'_4}\right|^2
\nonumber \\
&=&  \left(\frac{\alpha_s}{2\pi}\right)^2\;{\cal T}^0_{g g}(q^2)
\left(q^2 \right)^{-2\e} 
  \Bigg[ N_F^2 \Bigg( 
\frac{1}{9\e^2}  + \frac{7}{9\e} + \left(
  \frac{677}{162}
          - \frac{\pi^2}{6}\right) 
\Bigg) \nonumber \\
&& +  \frac{N_F}{N}
\Bigg( -\frac{5}{12} + \frac{\zeta_3}{3}
\Bigg)
+ {\cal O}(\e) \Bigg] \; .
\end{eqnarray}

The sum of all NNLO subtraction terms yields the following infrared
pole structure, which can be expressed in terms of NNLO infrared 
singularity operators~\cite{catani},
\begin{eqnarray}
\lefteqn{\Poles\left({\cal T}^2_{g gg}(q^2) + 
{\cal T}^2_{g q\bar q}(q^2) + {\cal T}^2_{g ggg}(q^2) 
+ {\cal T}^2_{g g q\bar q }(q^2)
+ {\cal T}^2_{q\bar q q'\bar q'}(q^2)\right) } \nonumber \\
&=& 
\left(\frac{\alpha_s}{2\pi}\right)^2\;{\cal T}^0_{g g}(q^2)
\left(q^2 \right)^{-2\e} 
  \Bigg[ N^2 \Bigg( -\frac{2}{\e^4} - \frac{121}{12\e^3} + \frac{1}{\e^2}
\left( -\frac{74}{9} + \frac{9\pi^2}{4} \right)
+ \frac{1}{\e} \left(
  \frac{65}{27}
          + \frac{33\pi^2}{8}
          + \frac{53}{6}\zeta_3 \right)  \Bigg) \nonumber \\
&& + N N_F \Bigg( 
\frac{11}{6\e^3}  + \frac{61}{18\e^2} + \frac{1}{\e} 
\left(-\frac{155}{108} - \frac{3\pi^2}{4} \right)
\Bigg) + \frac{N_F}{N} \Bigg( \frac{1}{4\e} 
\Bigg) 
+ N_F^2 \Bigg( - \frac{1}{3\e^2} \Bigg) 
+ {\cal O}(\e^0) \Bigg] \; 
\label{eq:polesrr}
\\
&=& - \left(\frac{\alpha_s}{2\pi}\right)^2\; \mbox{Re}\Bigg[ 
  - 2{\bom I}_{gg}^{(1)}(\e,q^2) 
\left(2{\bom I}_{gg}^{(1)}(\e,q^2) +2{\bom I}_{gg}^{(1),*}(\e,q^2)
 \right){\cal T}^0_{g g}(q^2)
  -2 \frac{\beta_0}{\e}  
\, 2{\bom I}_{gg}^{(1)}(\e,q^2) {\cal T}^0_{g g}(q^2)
 \nonumber\\
&&\hspace{2.4cm}
+ 4 \,  {\bom I}_{gg}^{(1)}(\e,q^2) {\cal T}^1_{g g}(q^2)
+ 2 \,
e^{-\e\gamma } \frac{ \Gamma(1-2\e)}{\Gamma(1-\e)} 
\left(\frac{\beta_0}{\e} + K\right)\, 2
 {\bom I}_{gg}^{(1)}(2\e,q^2) {\cal T}^0_{g g}(q^2) \nonumber\\
&& \hspace{2.4cm}
+ 2 \, {\bom H}_{g g}^{(2)}(\e,q^2){\cal T}^0_{g g}(q^2)
\, \Bigg] ,
\label{eq:I2}
\end{eqnarray}
where $\beta_0$ is the first term of the QCD $\beta$-function
(\ref{eq:qcdbeta}) and the constant $K$ 
\begin{equation}
K = \left( \frac{67}{18} - \frac{\pi^2}{6} \right) N - 
\frac{5}{9}  N_F.
\end{equation}
The final state dependent 
constant ${\bom H}_{g g}^{(2)}(\e,q^2)$ contributes only at 
${\cal O}(\e^{-1})$:
\begin{equation}
{\bom H}_{g g}^{(2)}(\epsilon,q^2)
=\frac{e^{\epsilon \gamma}}{4\,\epsilon\,\Gamma(1-\epsilon)} 
\,\left(2 H_g^{(2)}  \right)  \left(-q^2 \right)^{-2\e} \;.
\end{equation}
with
\begin{eqnarray}
H^{(2)}_g &=&  
\left(\frac{1}{2}\zeta_3+{\frac {5}{12}}+ {\frac {11\pi^2}{144}}
\right)N^2
+{\frac {5}{27}}\,\NF^2
+\left (-{\frac {{\pi }^{2}}{72}}-{\frac {89}{108}}\right ) N \NF 
-\frac{\NF}{4N}\;.
\label{eq:H2}
\end{eqnarray}

The above structure is to be compared with the renormalized purely virtual 
NNLO corrections (two-loop times tree plus 
one-loop self-interference), which were first computed by 
Harlander~\cite{harlander}:
\begin{eqnarray}
{\cal T}^2_{g g}(q^2) &\equiv& \frac{1}{2}\,
\left[2\mbox{Re}|{\cal M}^0_{g_1 g_2}{\cal M}^{2,*}_{g_1 g_2}
| + |{\cal M}^1_{g_1 g_2}|^2 \right]
\nonumber \\
&=&\left(\frac{\alpha_s}{2\pi}\right)^2\,
(q^2)^{-2\e}\, 
{\cal T}^0_{g g}(q^2) 
\Bigg[N^2 \Bigg( \frac{2}{\e^4} + \frac{121}{12\e^3} + \frac{1}{\e^2}
\left( \frac{74}{9} - \frac{9\pi^2}{4} 
\right)
+ \frac{1}{\e} \left(
  - \frac{65}{27}
          - \frac{33\pi^2}{8}
          - \frac{53}{6}\zeta_3 \right) 
\nonumber \\
&& \hspace{2cm}
+ \left(\frac{11369}{324} + \frac{335\pi^2}{72} - \frac{451}{18}\zeta_3
+ \frac{43\pi^4}{60} \right)
 \Bigg) \nonumber \\
&& \hspace{2cm} + N N_F \Bigg( -
\frac{11}{6\e^3}  - \frac{61}{18\e^2} + \frac{1}{\e} 
\left(\frac{155}{108} + \frac{3\pi^2}{4} \right)
+ \left(-\frac{6337}{648} - \frac{25\pi^2}{36} + \frac{23}{9}\zeta_3 \right)
\Bigg)\nonumber \\
&& \hspace{2cm}  + \frac{N_F}{N} \Bigg(- \frac{1}{4\e} 
+ \left(\frac{67}{24} - 2\zeta_3 \right)
\Bigg) 
+ N_F^2 \Bigg(  \frac{1}{3\e^2} \Bigg) 
+ {\cal O}(\e) \Bigg] \;.
\end{eqnarray}

It can be seen that the poles of the real radiation terms (\ref{eq:polesrr})
cancel the poles of the purely virtual corrections:
\begin{equation}
\Poles\left({\cal T}^2_{g gg}(q^2) + 
{\cal T}^2_{g q\bar q}(q^2) + {\cal T}^2_{g ggg}(q^2) 
+ {\cal T}^2_{g g q\bar q }(q^2)
+ {\cal T}^2_{q\bar q q'\bar q'}(q^2)\right)
+ \Poles\left({\cal T}^2_{gg}(q^2)\right) = 0 \;.
\end{equation}
The infrared singularity structure (\ref{eq:I2}) corresponds to the 
NNLO corrections to a tree level process 
containing {\it two} gluon-gluon antenna functions, 
as is the case for the Higgs boson decay. 

An important check on our results is that the sum of all NNLO contributions 
\begin{eqnarray}
\lefteqn{\Finite\left({\cal T}^2_{g gg}(q^2) + 
{\cal T}^2_{g q\bar q}(q^2) + {\cal T}^2_{g ggg}(q^2) 
+ {\cal T}^2_{g g q\bar q }(q^2)
+ {\cal T}^2_{q\bar q q'\bar q'}(q^2)\right) 
+ \Finite\left({\cal T}^2_{gg}(q^2) \right) } \nonumber \\
&=& \left(\frac{\alpha_s}{2\pi}\right)^2 \Bigg[N^2 \left(\frac{37631}{216} - \frac{121\pi^2}{36} - \frac{55}{2}\zeta_3
\right) + NN_F \left(-\frac{14509}{216} + \frac{11\pi^2}{9} + 2\zeta_3\right)
\nonumber \\ &&+\frac{N_F}{N} \left(\frac{131}{24} - 3\zeta_3 \right) 
+ N_F^2 \left( \frac{127}{27} - \frac{\pi^2}{9}\right)\, \Bigg] \, 
{\cal T}^0_{g g}(q^2)
\end{eqnarray}
agrees with the NNLO correction to the total Higgs 
boson decay rate into 
partons in the effective theory~\cite{kniehl}.

Equations~(\ref{eq:I2}) and (\ref{eq:H2}) demonstrate that 
the NNLO three and four parton contributions to Higgs boson decay into 
massless partons display the same singularity structure as 
final state observables containing adjacent gluon-gluon pairs. 
It is therefore possible 
 to derive colour-ordered gluon-gluon antenna functions 
from the Higgs boson decay matrix elements obtained here using the effective 
Lagrangian density (\ref{eq:lagr}).

\section{Conclusions and Outlook}
\setcounter{equation}{0}

QCD antenna functions describe the behaviour of QCD matrix elements in  their
infrared singular limits, corresponding to soft or collinear parton emission.
They are constructed so that they describe all singular limits arising  from
emission of unresolved partons in between the two colour-connected
hard partons that define the  antenna.  The
quark-antiquark antenna functions are directly related to the physical matrix
elements for $\gamma^* \to q\bar q +$partons.  We demonstrated in a 
previous paper~\cite{chi} that quark-gluon antenna functions could be obtained 
from an effective Lagrangian density describing neutralino decay into a
gluino and other partons. Besides quark-antiquark and quark-gluon antenna 
functions, QCD calculations also require gluon-gluon antenna functions. 
In this paper, we showed that gluon-gluon antenna functions can be 
obtained from physical Higgs boson decay matrix elements into partons, arising 
in an effective theory coupling the Higgs field to the gluonic field 
strength tensor.

We demonstrated that the physical Higgs boson decay matrix elements
reproduce the singular structure of QCD gluon-gluon antenna functions 
at NLO and NNLO. We extracted the infrared structure for decay kinematics, 
as required for jet observables without partons in the initial state. By 
analytic continuation, the matrix elements derived here can also be continued
to production (leading order process contains 
partons only in the initial state) or scattering (leading order 
process contains partons in initial and final state) kinematics, 
where they have to be integrated over the appropriate phase spaces.
The phase space integrals for production kinematics were derived 
in~\cite{higgsprod3}, such that the antenna subtraction terms 
for this kinematical situation can in principle be derived.

With this and two preceeding papers~\cite{our2j,chi}, we demonstrated that 
all QCD antenna functions can be derived 
(as opposed to constructed) from physical matrix elements: quark-antiquark 
antennae from the decay of a virtual photon into partons, quark-gluon antennae
from neutralino decay into gluino plus partons and finally gluon-gluon
antennae from Higgs boson decay into partons.
The NNLO antenna subtraction functions obtained through this procedure will be 
reported in  a subsequent publication~\cite{gggsub}.

\section*{Acknowledgements} 
This research was supported in part by the Swiss National Science Foundation 
(SNF) under contracts PMPD2-106101  and 200021-101874, 
 by the UK Particle Physics and Astronomy  Research Council and by
the EU Fifth Framework Programme  `Improving Human Potential', Research
Training Network `Particle Physics Phenomenology  at High Energy Colliders',
contract HPRN-CT-2000-00149.

\end{document}